# *Influence of Spatial Dispersion in the Topological Edge States of Magnetized Plasmas*

*João C. Serra*[1,*], *Mário G. Silveirinha*[1,†]

[(1)] *University of Lisbon–Instituto Superior Técnico and Instituto de Telecomunicações, Avenida Rovisco Pais, 1, 1049-001 Lisboa, Portugal*

**Abstract**

Conventional Chern insulators are two-dimensional periodic structures that support unidirectional edge states at the boundary, while the wave propagation in the bulk regions is forbidden. The number of unidirectional edge states is governed by the gap Chern number, a topological invariant that depends on the global properties of the system over the entire wavevector space. This concept can also be extended to systems with a continuous translational symmetry provided they satisfy a regularization condition for large wavenumbers. Here, we discuss how the spatial dispersion, notably the high-spatial frequency behavior of the material response, critically influences the topological properties, and consequently, the net number of unidirectional edge states. In particular, we show that seemingly small perturbations of a local magnetized plasma can lead to distinct Chern phases and, consequently, markedly different edge state dispersions.

---

* E-mail: joao.serra@lx.it.pt

† To whom correspondence should be addressed: E-mail: mario.silveirinha@tecnico.ulisboa.pt

# I. Introduction

Lorentz's reciprocity theorem states that the field produced by a given source remains the same if we interchange the source and observation points in a linear time-reversal symmetric (Hermitian) system [1, 2]. Naturally, such systems cannot support asymmetric light flows, and thus cannot provide optical isolation. To surpass this constraint, different methods have been proposed to design nonreciprocal devices by breaking time-reversal symmetry via an external magnetic field bias [3-5], moving media [6-11] and time-varying modulations [12-15]. Alternatively, nonreciprocity can also be achieved by exploiting non-Hermitian [16-20] and nonlinear effects [21-24].

Within this context, topological photonics has emerged as a unique paradigm to realize unidirectional edge-type waveguides robust to path deformations and other irregularities. Chern insulators first appeared in condensed matter theory in connection to the discovery of the quantum Hall effect [25-27], but since then they have transitioned to photonics [28-30] and other fields [31-33].

Suppose that we have a physical system whose spectrum is determined by the eigenvalue problem $\hat{H}(\mathbf{k})\cdot\boldsymbol{\psi}_{n,\mathbf{k}} = \omega_{n,\mathbf{k}}\boldsymbol{\psi}_{n,\mathbf{k}}$ with $\mathbf{k}$ some vector defined over a two-dimensional closed (i.e., compact and without boundary) manifold $S$. If the operator $\hat{H}(\mathbf{k})$ varies smoothly with respect to $\mathbf{k}$ in the entire manifold $S$, we can assign an integer gap Chern number $\mathcal{C}_{\text{gap}}$ to each band gap, i.e., a spectral region without eigenvalues $\omega_{n,\mathbf{k}}$ [34-36]. In periodic systems, $\mathbf{k} = (k_x, k_y)$ usually denotes the wavevector of a Bloch wave, whereas the eigenvalue $\omega_{n,\mathbf{k}}$ determines the frequency of the time-harmonic modes $\psi_{n,\mathbf{k}} \sim e^{-i\omega_{n,\mathbf{k}}t}$. Similar to the genus $g$ of a closed surface, this topological invariant is a property of the system as a whole and it is robust to perturbations that do not close the frequency gap [37].

Far from being a mere mathematical curiosity, the Chern topology exhibits significant physical implications: in electronic systems, it determines the quantized Hall conductivity, whereas in photonics it governs the fluctuation-induced light-angular momentum of an insulator cavity [38, 39]. Moreover, the gap Chern number determines the net number of unidirectional edge modes propagating at the interface between a topological material and a trivial insulator [40]. This relation between the bulk properties of a material and the edge dispersion is known as the bulk-edge correspondence.

Conventional topological insulators are two-dimensional periodic structures (crystals), where the wavevector manifold $S$ is a Brillouin zone homeomorphic to a torus [41-47]. Nevertheless, other base manifolds may be considered as well. For example, in continuous media, we can "compactify" the (unbounded) Euclidean plane $\mathbb{R}^2$ into a Riemann sphere. However, if the material response does not satisfy certain regularization conditions for large wavenumbers, this compactification does not lead to an operator $\hat{H}(\mathbf{k})$ that is well-behaved over a closed manifold and the Chern theorem no longer applies [48-50].

In such a case, the system topology is ill-defined, prompting several questions: What happens if the gap Chern number is non-integer? How can the material response be adjusted to guarantee the emergence of topological phases? Is the regularized topology unique, or does it depend on the regularization process? Some of these questions have been partially explored in previous works [48, 50-55], which revealed intriguing effects, such as energy sinks [56, 57], in platforms with ill-defined topologies. Moreover, ill-defined Chern insulators are not exclusive of continuous media, they also exist in photonic crystals characterized by a frequency-dependent material response [58]. In this article, we present for the first time a comprehensive study of the topology and edge states of magnetized plasmas with various types of dispersive responses. In particular, we demonstrate how spatial dispersion can critically influence the topology of a magnetized plasma and the number of edge states supported within a topological band gap.

The paper is organized as follows. In section II, we present a brief review of topological concepts and of the Chern theorem in continuous media. In Sect. III, we characterize the Chern phases for various physical models of a magnetized plasma. In section IV, we demonstrate how these seemingly similar models result in distinct edge-state dispersions due to the bulk-edge correspondence. In section V, we present a detailed comparison between the different models and a summary of the main results.

## II. Chern Theorem in Continuous Media

Consider the spectral problem

$$\hat{H}(\mathbf{k})\cdot\boldsymbol{\psi}_{n,\mathbf{k}} = \omega_{n,\mathbf{k}}\boldsymbol{\psi}_{n,\mathbf{k}}, \tag{1}$$

where the wavevector $\mathbf{k}=(k_x,k_y)$ is defined over a two-dimensional manifold $S$. For each separable band $\omega_{n,\mathbf{k}}$, we can assign a Chern number

$$\mathcal{C}_n = \frac{1}{2\pi}\iint_S d\mathbf{k}\ \mathcal{F}_{n,\mathbf{k}} \tag{2}$$

with the Berry curvature defined as

$$\mathcal{F}_{n,\mathbf{k}} = \hat{\mathbf{z}} \cdot \left( \nabla_{\mathbf{k}} \times \mathbf{A}_{n,\mathbf{k}} \right), \quad \mathbf{A}_{n,\mathbf{k}} = \mathrm{Re}\left\{ i \boldsymbol{\psi}_{n,\mathbf{k}}^{*} \cdot \nabla_{\mathbf{k}} \boldsymbol{\psi}_{n,\mathbf{k}} \right\}. \tag{3}$$

The eigenvectors $\boldsymbol{\psi}_{n,\mathbf{k}}$ are assumed orthonormal, i.e., $\left\langle \boldsymbol{\psi}_{n,\mathbf{k}}, \boldsymbol{\psi}_{m,\mathbf{k}} \right\rangle = \delta_{n,m}$, where $\left\langle \cdot, \cdot \right\rangle$ denotes an inner product with respect to which the operator $\hat{H}\left(\mathbf{k}\right)$ is Hermitian. In general, $\mathcal{C}_n$ may take any real value. However, if the base manifold $S$ is closed and the operator $\hat{H}\left(\mathbf{k}\right)$ is smooth, then the Chern theorem establishes that $\mathcal{C}_n$ is necessarily an integer and it is invariant to small perturbations that preserve the separability of the considered band ($\omega_{n,\mathbf{k}}$).

As explained in the Introduction, the wavevector space in continuous media is the unbounded Euclidean plane $\mathbb{R}^2$ and thus some caution is required to compute topological Chern invariants in these platforms. To circumvent this obstacle, Ref. [48] suggests compactifying the base manifold by adding a single "point at infinity", thus becoming isomorphic to the Riemann sphere. In that case, we can use Stoke's theorem to compute the Chern number as

$$\mathcal{C}_n = \frac{1}{2\pi} \sum_m \oint_{C(\mathbf{k}_m)} d\mathbf{k} \cdot \mathbf{A}_{n,\mathbf{k}} + \frac{1}{2\pi} \oint_{C(\infty)} d\mathbf{k} \cdot \mathbf{A}_{n,\mathbf{k}}. \tag{4}$$

Above, we denote $C\left(\mathbf{k}_m\right)$ as a circumference of arbitrarily small radius around a singularity of the (gauge-dependent) Berry potential $\mathbf{A}_{n,\mathbf{k}}$ located at $\mathbf{k} = \mathbf{k}_m \in \mathbb{R}^2$ and $C\left(\infty\right)$ is a closed path of arbitrarily large radius to account for the contributions at $\mathbf{k} = \infty$. In rotational-invariant problems (such as all the examples considered in this article), it is usually possible to choose an eigenvector basis that is rotationally symmetric and smooth everywhere except perhaps at the origin $\mathbf{k} = \mathbf{0}$ and at infinity $\mathbf{k} = \infty$:

$$\mathcal{C}_n = \frac{1}{2\pi} \oint_{C(\infty)} d\mathbf{k} \cdot \mathbf{A}_{n,\mathbf{k}} - \frac{1}{2\pi} \oint_{C(\mathbf{0})} d\mathbf{k} \cdot \mathbf{A}_{n,\mathbf{k}}. \quad (5)$$

Naturally, the Chern theorem only holds if the compactification leads to a sufficiently well-behaved operator $\hat{H}(\mathbf{k})$ (see Fig. 1). For that, the material response must satisfy a suitable regularization condition as $|\mathbf{k}| \to \infty$. Otherwise, the Chern number may be non-integer, and the system is topologically ill-defined.

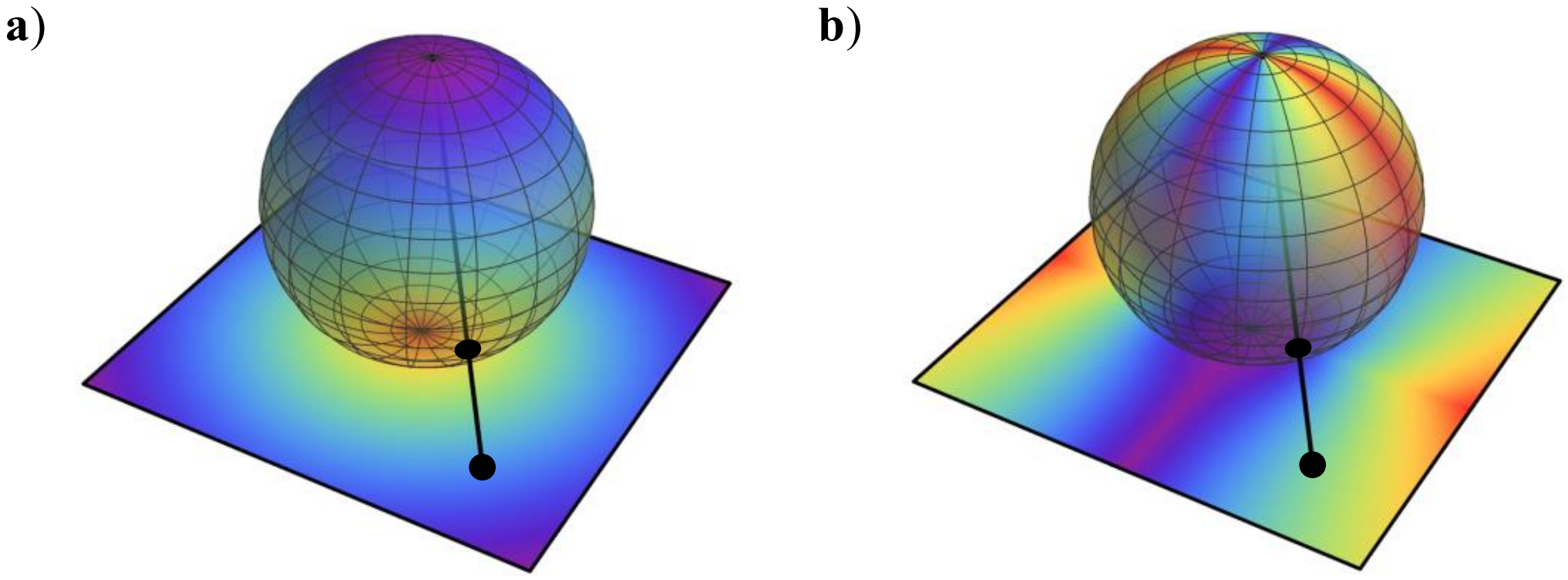

**Fig. 1** Visual representation of how the compactification of the unbounded Euclidean plane $\mathbb{R}^2$ creates well-behaved **(a)** or ill-behaved **(b)** operators near $\mathbf{k} = \infty$. The color map represents the Berry curvature. The points on the compactified plane are mapped one-to-one onto the points of the Riemann sphere using the stereographic projection.

## III. Magnetized Electric Plasma

In order to illustrate how the high-spatial frequency response of a material ($|\mathbf{k}| \to \infty$) can critically impact its topology, next we consider different physical descriptions of an electron gas subject to an external magnetic field bias $\mathbf{B}_0 = B_0\hat{\mathbf{z}}$ as illustrated in Fig. 2a. This material is known to exhibit nonreciprocal phenomena, such as the Faraday rotation effect [3], because the magnetically-induced cyclotron orbits of the free electrons break time-reversal symmetry. The medium effectively behaves as a continuum of rotating wheels, generating a rotational drag that underlies the Faraday effect [59, 60]. For simplicity, we restrict the propagation of electromagnetic waves to the $xoy$ plane ( $\partial / \partial z = 0$ ) and assume a transverse-magnetic (TM) polarization, i.e., $\mathbf{B} \sim \hat{\mathbf{z}}$, to exploit this nonreciprocal response.

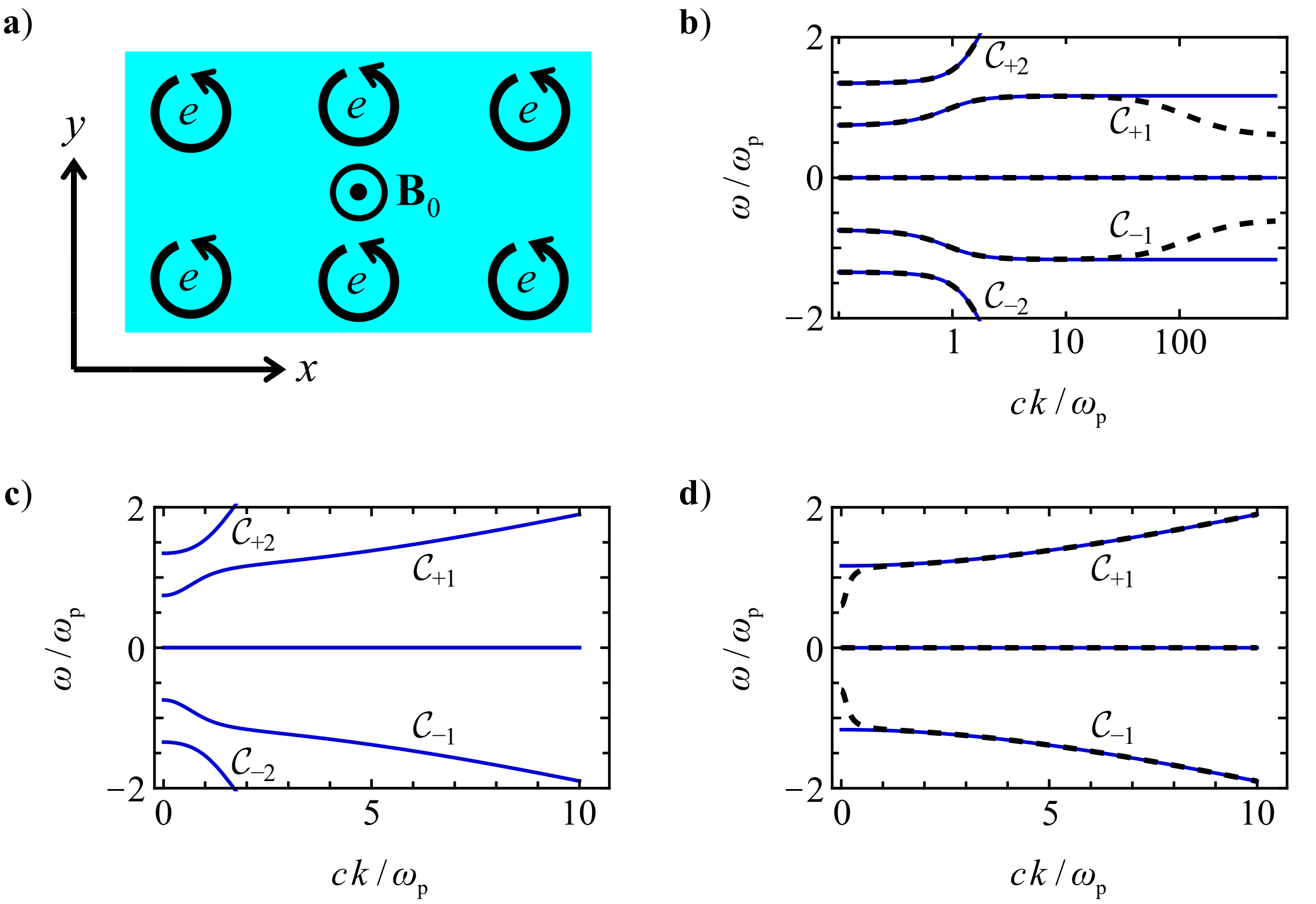

**Fig. 2** **(a)** Visual representation of the electron cyclotron orbits in a magnetized plasma. **(b)-(d)** Band structures of TM modes in a magnetized plasma described by different physical models. The cyclotron frequency is $\omega_c = 0.6\omega_p$. **(b)** Local model (dashed black) and full-cutoff model (solid blue) with a large-wavevector cutoff $k_{max} = 100\omega_p / c$. The Chern numbers are $\mathcal{C}_{\pm 2} = \mp 1$ and $\mathcal{C}_{\pm 1} \approx \pm 1.514$ for the local model, and $\mathcal{C}_{\pm 2} = \mp 1$ and $\mathcal{C}_{\pm 1} = \pm 2$ for the full-cutoff model. **(c)** Hydrodynamic model for a diffusion velocity $\beta = 0.15c$: $\mathcal{C}_{\pm 2} = \mp 1$ and $\mathcal{C}_{\pm 1} = \pm 1$. **(d)** Quasi-static approximation of the hydrodynamic model

with (dashed black) and without (solid blue) a low-wavevector cutoff $k_{\min} = 0.2\omega_{\mathrm{p}} / c$. The Chern numbers are $\mathcal{C}_{\pm 1} \approx \mp 0.514$ for the unregularized model and $\mathcal{C}_{\pm 1} = \mp 1$ for the regularized model.

### *A. Local and Full-cutoff Models*

We begin by considering a local model that results from combining the microscopic Maxwell's equations,

$$-\nabla \times \mathbf{E} = \partial_t \mathbf{B}, \tag{6a}$$

$$+\nabla \times \mathbf{B} - \mu_0 \mathbf{J} = \varepsilon_0 \mu_0 \partial_t \mathbf{E} \tag{6b}$$

with an electron transport equation in the absence of dissipative collisions,

$$nq\mathbf{E} - \mathbf{B}_0 \times \mathbf{J} = \frac{m}{q} \partial_t \mathbf{J}. \tag{7}$$

Above, $\mathbf{E}$ and $\mathbf{B}$ denote the dynamic electric and the magnetic fields, respectively, $\mathbf{J}$ is the electric current density, $n$ is the average electron density, $m$ is the electron's rest mass, and $q = -e$ is the electron's charge. Note that the left-hand side of Eq. (7) corresponds to the Lorentz force per unit of volume. For simplicity, we assume $|\mathbf{B}| << |B_0|$.

Since the system is invariant under continuous space and time translations, we look for plane wave solutions ($\nabla \to +i\mathbf{k}$, $\partial_t \to -i\omega$). This leads to the conventional Hermitian eigenvalue problem:

$$H_{\mathrm{local}}(\mathbf{k}) \cdot \boldsymbol{\psi}_{n,\mathbf{k}} = \frac{\omega_{n,\mathbf{k}}}{\omega_{\mathrm{p}}} \boldsymbol{\psi}_{n,\mathbf{k}} \tag{8}$$

with

$$H_{\text{local}}(\mathbf{k}) = \begin{pmatrix} \mathbf{0} & -\frac{c\mathbf{k}}{\omega_{\text{p}}}\times\mathbf{1} & -i\mathbf{1} \\ +\frac{c\mathbf{k}}{\omega_{\text{p}}}\times\mathbf{1} & \mathbf{0} & \mathbf{0} \\ +i\mathbf{1} & \mathbf{0} & +i\frac{\omega_{\text{c}}}{\omega_{\text{p}}}\hat{\mathbf{z}}\times\mathbf{1} \end{pmatrix}, \quad \psi_{n,\mathbf{k}} = \begin{pmatrix} \sqrt{\varepsilon_0}\mathbf{E}_{n,\mathbf{k}} \\ \mathbf{B}_{n,\mathbf{k}}/\sqrt{\mu_0} \\ \mathbf{J}_{n,\mathbf{k}}/\omega_{\text{p}}\sqrt{\varepsilon_0} \end{pmatrix}. \tag{9}$$

We have introduced the bulk plasma frequency $\omega_{\text{p}} = \sqrt{\frac{nq^2}{\varepsilon_0 m}}$ and the cyclotron frequency $\omega_{\text{c}} = -\frac{q}{m}B_0$. It is implicit that $\mathbf{E}_{n,\mathbf{k}} = E_x\hat{\mathbf{x}} + E_y\hat{\mathbf{y}}$, $\mathbf{J}_{n,\mathbf{k}} = J_x\hat{\mathbf{x}} + J_y\hat{\mathbf{y}}$ and $\mathbf{B}_{n,\mathbf{k}} = B_z\hat{\mathbf{z}}$, so that the Hamiltonian is represented by a 5×5 matrix.

The eigenvalue problem (8) has five TM-polarized frequency bands as illustrated in Fig. 2b. In agreement with previous works [61, 62], we show in Appendix A that the high-frequency bands $\omega_{\pm 2,\mathbf{k}}$ are characterized by integer Chern numbers $\mathcal{C}_{\pm 2} = \mp \text{sgn}[\omega_{\text{c}}]$, while the low-frequency bands $\omega_{\pm 1,\mathbf{k}}$ have non-integer topological invariants $\mathcal{C}_{\pm 1} = \pm \text{sgn}[\omega_{\text{c}}]\left(1 + 1/\sqrt{1 + \frac{\omega_{\text{p}}^2}{\omega_{\text{c}}^2}}\right)$. The particle-hole symmetry implies that the Chern number of a positive frequency band is the symmetric of the corresponding negative frequency band. Therefore, throughout this work, we ignore the static bands ($\omega_{\mathbf{k}} = 0$) as their total topological charge is always trivial.

To understand the origin of this ill-defined topology, it is convenient to consider the equivalent macroscopic (relative) permittivity tensor:

$$\bar{\varepsilon}_{\text{local}}(\omega) = \varepsilon_t(\omega)\mathbf{1}_t + i\varepsilon_g(\omega)\hat{\mathbf{z}}\times\mathbf{1} + \varepsilon_z(\omega)\hat{\mathbf{z}}\otimes\hat{\mathbf{z}},$$
$$\varepsilon_t(\omega) = 1 - \frac{\omega_{\text{p}}^2}{\omega^2 - \omega_{\text{c}}^2}, \quad \varepsilon_g(\omega) = -\frac{\omega_{\text{c}}}{\omega}\frac{\omega_{\text{p}}^2}{\omega^2 - \omega_{\text{c}}^2}, \quad \varepsilon_z(\omega) = 1 - \frac{\omega_{\text{p}}^2}{\omega^2}. \tag{10}$$

The material response for the eigenmodes $\boldsymbol{\psi}_{\pm 2,\mathbf{k}}$ becomes trivial for large wavenumbers, i.e., $\lim_{|\mathbf{k}|\to\infty} \bar{\varepsilon}_{\text{local}}(\omega_{\pm 2,\mathbf{k}}) = \mathbf{1}$, so that the response becomes reciprocal for $|\mathbf{k}|\to\infty$. In that case, it is possible to pick a gauge such that the Berry potential $\mathbf{A}_{\pm 2,\mathbf{k}}$ vanishes as $|\mathbf{k}|\to\infty$, implying $\mathcal{C}_{\pm 2}\in\mathbb{Z}$ [48]. On the contrary, eigenmodes $\boldsymbol{\psi}_{\pm 1,\mathbf{k}}$ with large wavenumbers experience a nonreciprocal electric response characterized by the gyrotropic parameter $\lim_{|\mathbf{k}|\to\infty} \varepsilon_g(\omega_{\pm 1,\mathbf{k}}) = -\frac{\omega_{\text{c}}}{\omega_{\text{p}}^2 + \omega_{\text{c}}^2}$. Thus, in the latter case the vector potential has an intrinsic singularity (which cannot be removed with a gauge transformation) near $\mathbf{k}=\infty$ (North Pole of the Riemann sphere). This explains the ill-defined topology of the local model [48].

From a physical perspective, every material should asymptotically behave as the vacuum for $\mathbf{k}\to\infty$, because an electric dipole with length $l$ does not react to excitations with very short wavelengths $\lambda = 2\pi/|\mathbf{k}| << l$ [63]. Therefore, one way to regularize the topology of the local model is to introduce a large-wavevector cutoff $k_{\max}$ that enforces this physical constraint [48], i.e.,

$$\bar{\varepsilon}_{\text{cutoff}}(\omega,\mathbf{k}) = \mathbf{1} + \frac{1}{1 + k^2/k_{\max}^2}\left[\bar{\varepsilon}_{\text{local}}(\omega) - \mathbf{1}\right] \tag{11}$$

with $k^2 = \mathbf{k} \cdot \mathbf{k}$. For $k << k_{\text{max}}$, we retrieve the local model as $\bar{\varepsilon}_{\text{cutoff}}(\omega, \mathbf{k}) \approx \bar{\varepsilon}_{\text{local}}(\omega)$, whereas for large wavenumbers $k >> k_{\text{max}}$ the response becomes trivial, i.e., $\bar{\varepsilon}_{\text{local}}^{\text{cutoff}}(\omega, \mathbf{k}) \approx \mathbf{1}$.

This full-cutoff model is described by the regularized Hamiltonian:

$$H_{\text{cutoff}}(\mathbf{k}) = \begin{pmatrix} \mathbf{0} & -\dfrac{c\mathbf{k}}{\omega_{\text{p}}} \times \mathbf{1} & -i\sqrt{\dfrac{1}{1+k^2/k_{\text{max}}^2}}\mathbf{1} \\ +\dfrac{c\mathbf{k}}{\omega_{\text{p}}} \times \mathbf{1} & \mathbf{0} & \mathbf{0} \\ +i\sqrt{\dfrac{1}{1+k^2/k_{\text{max}}^2}}\mathbf{1} & \mathbf{0} & +i\dfrac{\omega_{\text{c}}}{\omega_{\text{p}}}\hat{\mathbf{z}} \times \mathbf{1} \end{pmatrix}. \qquad (12)$$

The corresponding dispersion for the bulk bands is represented in Fig. 2b with a dashed black curve. As seen, it is very similar to the one of the local model (solid blue curve) for $k << k_{\text{max}}$. Now, the Chern topology is regularized so that $\mathcal{C}_{\pm 1} = \pm 2\,\text{sgn}[\omega_{\text{c}}]$ and $\mathcal{C}_{\pm 2} = \mp \text{sgn}[\omega_{\text{c}}]$ (see Appendix A). Importantly, this result is independent of the specific value of wavevector cutoff, provided $0 < k_{\text{max}} < \infty$.

*B. Hydrodynamic Model*

Next, we consider the well-known hydrodynamic (drift-diffusion) model for the magnetized plasma that accounts for the electron-electron repulsive interactions [54, 64-67]. The corresponding transport equation is of the form:

$$+\varepsilon_0 \omega_{\text{p}}^2 \mathbf{E} + \omega_{\text{c}} \hat{\mathbf{z}} \times \mathbf{J} - \beta^2 \nabla \rho = \partial_t \mathbf{J}, \qquad (13)$$

where $\rho$ denotes the free-electron charge density, which is linked to the current density through the continuity equation:

$$\nabla \cdot \mathbf{J} + \partial_t \rho = 0. \tag{14}$$

The last term in the left-hand side of the transport equation models the effects of diffusion. The parameter $\beta$ has unities of velocity and is proportional to the Fermi velocity of the metal. The spectral problem $H_{\text{hydro}}(\mathbf{k}) \cdot \boldsymbol{\psi}_{n,\mathbf{k}} = \frac{\omega_{n,\mathbf{k}}}{\omega_{\text{p}}} \boldsymbol{\psi}_{n,\mathbf{k}}$ associated with the hydrodynamic model is governed by the following matrix operator:

$$H_{\text{hydro}}(\mathbf{k}) = \begin{pmatrix} \mathbf{0} & -\frac{c\mathbf{k}}{\omega_{\text{p}}} \times \mathbf{1} & 0 & -i\mathbf{1} \\ +\frac{c\mathbf{k}}{\omega_{\text{p}}} \times \mathbf{1} & \mathbf{0} & 0 & \mathbf{0} \\ \mathbf{0} & \mathbf{0} & 0 & +\frac{\beta\mathbf{k}}{\omega_{\text{p}}} \cdot \\ +i\mathbf{1} & \mathbf{0} & +\frac{\beta\mathbf{k}}{\omega_{\text{p}}} & +i\frac{\omega_{\text{c}}}{\omega_{\text{p}}} \hat{\mathbf{z}} \times \mathbf{1} \end{pmatrix}, \tag{15a}$$

with the state vector defined as:

$$\boldsymbol{\psi}_{n,\mathbf{k}} = \left( \sqrt{\varepsilon_0}\mathbf{E}, \quad \frac{\mathbf{B}}{\sqrt{\mu_0}}, \quad \frac{\beta\rho}{\omega_{\text{p}}\sqrt{\varepsilon_0}}, \quad \frac{\mathbf{J}}{\omega_{\text{p}}\sqrt{\varepsilon_0}} \right)^T. \tag{15b}$$

As before, it is implicit that $\mathbf{E}_{n,\mathbf{k}} = E_x\hat{\mathbf{x}} + E_y\hat{\mathbf{y}}$, $\mathbf{J}_{n,\mathbf{k}} = J_x\hat{\mathbf{x}} + J_y\hat{\mathbf{y}}$ and $\mathbf{B}_{n,\mathbf{k}} = B_z\hat{\mathbf{z}}$, so that the operator is represented by a $6 \times 6$ matrix. As illustrated in Fig. 2c, the band structure consists of two positive frequency bands, two negative frequency bands and two static-like bands. As the diffusion effects prevent wave localization on spatial scales smaller than $L \sim \beta / \omega_p$, the photonic dispersion does not saturate for large wavevectors, i.e., $\omega_{\pm 1,\mathbf{k}}^2 \sim \beta^2 k^2$ and $\omega_{\pm 2,\mathbf{k}}^2 \sim c^2 k^2$ as $k \to \infty$. As a result, Appendix A shows that the Chern numbers of these bands are integer numbers: $\mathcal{C}_{\pm 1} = \pm \text{sgn}[\omega_{\text{c}}]$ and $\mathcal{C}_{\pm 2} = \mp \text{sgn}[\omega_{\text{c}}]$ [54]. In fact, it is possible to prove, using perturbation theory, that a nonreciprocal perturbation of a reciprocal system always leads to integer Chern numbers for a band such that $\omega_{n,\mathbf{k}} \to \infty$, provided that $\omega_{n,\mathbf{k}} - \omega_{m,\mathbf{k}} \to \infty$ for all the other bands $m \neq n$, as happens in the hydrodynamic model for all *n*.

It is important to underline that both the full-cutoff model and the hydrodynamic model yield the same band structure as the local model when $1/k_{\max} = 0$ and $\beta = 0$, respectively. Accordingly, the Chern numbers of the high-frequency band in the two regularized models are identical. Interestingly, despite this property, the Chern topology of the low-frequency band in the hydrodynamic model differs from that of the full-cutoff model. This difference arises because the mathematical convergence associated with the limits $1/k_{\max} \to 0$ and $\beta \to 0$ is not uniform. In other words, the asymptotic behavior of the modes near $k = \infty$ remains quite different in the two models, even when the respective cutoff parameters ($1/k_{\max}$ or $\beta$) are arbitrarily small, resulting in two distinct topologies.

In fact, topological features are global properties and thereby can be greatly impacted by the high-spatial frequency response of a material, specifically by spatial dispersion effects. The local model topology is ill-defined because its high-spatial frequency response has discontinuous features that prevent a proper topological classification.

### *C. Quasi-Static Limit*

Quasi-static models are frequently used in plasmonics, as they significantly simplify the analysis of electromagnetic processes in the near-field region, where retardation effects are negligible. Within this framework, the dynamical magnetic field can be neglected ($\mathbf{B} \approx \mathbf{0}$) and the electric field is written in terms of a scalar electric potential, $\mathbf{E} \approx -\nabla \phi$, governed by Gauss' law,

$$\nabla \cdot \mathbf{E} = \frac{\rho}{\varepsilon_0} \Leftrightarrow \phi = \frac{\rho}{\varepsilon_0 k^2}. \tag{16}$$

Next, we employ this quasi-static approximation to the hydrodynamic model of the previous subsection. In this context, the quasi-static approximation is expected to be applicable for waves such that $\omega / c << k$, which are dominant in the near-field region.

Combining Gauss's law [Eq. (16)] with the transport and continuity equations [Eqs. (13)-(14)], we find that the quasi-static dynamics is modelled by the eigenvalue problem $H_{\text{QS}}(\mathbf{k}) \cdot \boldsymbol{\psi}_{n,\mathbf{k}} = \frac{\omega_{n,\mathbf{k}}}{\omega_{\text{p}}} \boldsymbol{\psi}_{n,\mathbf{k}}$ with

$$H_{\text{QS}}(\mathbf{k}) = \begin{pmatrix} 0 & +\sqrt{\beta^2 + \frac{\omega_{\text{p}}^2}{k^2}} \frac{\mathbf{k}}{\omega_{\text{p}}} \cdot \\ +\sqrt{\beta^2 + \frac{\omega_{\text{p}}^2}{k^2}} \frac{\mathbf{k}}{\omega_{\text{p}}} & +i \frac{\omega_{\text{c}}}{\omega_{\text{p}}} \hat{\mathbf{z}} \times \mathbf{1} \end{pmatrix}, \quad \boldsymbol{\psi}_{n,\mathbf{k}} = \begin{pmatrix} \sqrt{\beta^2 + \frac{\omega_{\text{p}}^2}{k^2}} \frac{\rho_{n,\mathbf{k}}}{\omega_{\text{p}} \sqrt{\varepsilon_0}} \\ \frac{\mathbf{J}_{n,\mathbf{k}}}{\omega_{\text{p}} \sqrt{\varepsilon_0}} \end{pmatrix}. \tag{17}$$

As expected, this approximation only captures the lower-frequency bands of the hydrodynamic model (see Fig. 2d), which are associated with longitudinal-like volume plasmon modes. As the single positive frequency band satisfies $\omega_{+1,\mathbf{k}} \to +\infty$, one might expect a well-defined topology. However, it turns out that the Chern number is once again ill-defined (see Appendix A): $\mathcal{C}_{\pm 1} = \mp \text{sgn}[\omega_{\text{c}}] \left( 1 / \sqrt{1 + \frac{\omega_{\text{p}}^2}{\omega_{\text{c}}^2}} \right)$. Unlike the local model, this is unrelated to the non-compactness of the wavevector space. Instead, it stems from the discontinuous behavior of the operator $H_{\mathrm{QS}}(\mathbf{k})$ at the origin $\mathbf{k}=\mathbf{0}$, due to the elements proportional to $\sqrt{\beta^2+\frac{\omega_{\mathrm{p}}^2}{k^2}}\frac{\mathbf{k}}{\omega_{\mathrm{p}}}$.

The singularity should be understood as a deficiency of this simplified model because the quasi-static approximation is only valid for large wavenumbers. The problem can be fixed in different ways. The first one is to ignore the effects of the electric force in the transport equation which corresponds to the limit $\omega_{\mathrm{p}}\to 0$. This is the basis of the analysis of Ref. [65]. Alternatively, one can introduce a low-spatial frequency cutoff $k_{\min}$ in Eq. (16), i.e.,

$$\phi=\frac{\rho}{\varepsilon_0\left(k^2+k_{\min}^2\right)}, \tag{18}$$

to obtain a smooth operator near the origin:

$$H_{\mathrm{QS}}^{\mathrm{cutoff}}(\mathbf{k})=\begin{pmatrix} 0 & +\sqrt{\beta^2+\frac{\omega_{\mathrm{p}}^2}{k^2+k_{\min}^2}}\frac{\mathbf{k}}{\omega_{\mathrm{p}}}\cdot \\ +\sqrt{\beta^2+\frac{\omega_{\mathrm{p}}^2}{k^2+k_{\min}^2}}\frac{\mathbf{k}}{\omega_{\mathrm{p}}} & +i\frac{\omega_{\mathrm{c}}}{\omega_{\mathrm{p}}}\hat{\mathbf{z}}\times\mathbf{1} \end{pmatrix}. \tag{19}$$

Evidently, in the limit $k_{\min}\to 0$ we recover the original quasi-static model. The resulting bulk dispersion is shown in Fig. 2d and the Chern numbers are now integer numbers: $\mathcal{C}_{\pm 1}=\mp\operatorname{sgn}[\omega_{\mathrm{c}}]$. Curiously, these Chern numbers do not coincide with those of the hydrodynamic model ($\mathcal{C}_{\pm 1}=\pm\operatorname{sgn}[\omega_{\mathrm{c}}]$). This highlights once again that Chern topology is a global property, and approximations that neglect the detailed behavior of the system's response in certain regions of the parametric space (in this case, near the origin) can lead to different topologies.

It is worth mentioning that other methods can be employed to regularize the response of a continuous system, and these may lead to different classes of Chern insulators [54, 65, 68-70].

## IV. Edge States Dispersion

One of the most notable features of Chern insulators is the bulk-edge correspondence. The walls of a photonic insulator cavity characterized by the gap Chern number $\mathcal{C}_{\text{gap}}$ support $N_+$ counter-clockwise (CCW) propagating edge states and $N_- = N_+ + \mathcal{C}_{\text{gap}}$ clockwise (CW) edge states across the entire band gap [40]. The gap Chern number is defined as

$$\mathcal{C}_{\text{gap}} = \sum_{\omega_{n,\mathbf{k}} < \omega_{\text{gap}}} \mathcal{C}_n, \tag{20}$$

where the sum is performed over all the frequency bands $\omega_{n,\mathbf{k}}$ below the gap. Moreover, the topological nature of the Chern invariants guarantees that these edge-type channels are robust against perturbations that do not close the band gap. The gap Chern number can also be directly computed using a Green's function formalism without the need to determine all the eigenmodes of the system [55, 62].

In the previous section, we demonstrated that seemingly similar physical descriptions of a magnetized plasma originate very different Chern topologies, regardless of how small the parameter controlling the nonlocal perturbation may be. Next, we show how these inequivalent Chern insulators shape the dispersion of edge modes in distinct ways, consistently aligning with the bulk-edge correspondence. We compare the propagation properties of the topological edge states with the dispersion of the edge states supported by the local model, which is associated with an ill-defined topology with non-integer Chern numbers.

To explore this, let us consider a flat interface ($y=0$) that separates the half-region $y>0$ filled with a bulk magnetized plasma, from the other half-region $y<0$ filled with a perfect electric conductor (PEC) as illustrated in Fig. 3a. In this planar geometry, the CW/CCW edge states of the cavity geometry correspond to the edge states propagating along the $-x/+x$ direction, respectively, as represented by the red/blue squiggly arrow.

We look for TM edge states in the electron gas region ($y>0$) described by linear combinations of evanescent time-harmonic plane waves ($e^{-i\omega t}$) with a wavevector of the form $\mathbf{k}_n = q\hat{\mathbf{x}} + i\gamma_n\hat{\mathbf{y}}$, corresponding to the spatial variation $e^{i\mathbf{k}_n\cdot\mathbf{r}} = e^{iqx}e^{-\gamma_n y}$. The allowed wavevectors are determined by the characteristic equation:

$$\det\left( \mathbf{H}(\mathbf{k})\Big|_{\mathbf{k}=q\hat{\mathbf{x}}+i\gamma\hat{\mathbf{y}}} - \frac{\omega}{\omega_{\mathrm{p}}}\mathbf{1} \right) = 0. \tag{21}$$

The matrix operator $\mathbf{H}(\mathbf{k})$ depends on the physical model of the plasma. The characteristic equation is solved with respect to the attenuation constant $\gamma$, subject to the constraint $\mathrm{Re}\{\gamma\}>0$ to ensure that the energy is localised at the interface $y=0^+$. As it will be further discussed below, in general there are multiple solutions. The number of solutions is denoted by *N*.

The system state vector in the electron gas region is written in terms of a linear combination of the allowed evanescent plane-wave modes:

$$\boldsymbol{\psi} = e^{iqx} e^{-i\omega t} \sum_{n=1}^{N} \alpha_n \boldsymbol{\psi}_{n,\mathbf{k}} e^{-\gamma_n y} , \quad y > 0 . \tag{22}$$

Here, $\boldsymbol{\psi}_{n,\mathbf{k}}$ denotes the envelope of a generic plane wave and is defined as in Appendix A for each of the physical models, with the replacements $\omega_{n,\mathbf{k}} \to \omega$ and $\mathbf{k} \to q\hat{\mathbf{x}} + i\gamma_n \hat{\mathbf{y}}$. The complex coefficients $\alpha_n$ determine the weights of the linear combination. As further discussed in Appendix B and in the following subsections, the dispersion of the edge states is obtained by enforcing suitable boundary conditions.

*A. Local and Full-cutoff Models*

For the local model [Eq. (8)], the characteristic equation (21) has a single solution ( $N = 1$). The edge states dispersion is obtained by enforcing $E_x = 0$ at the boundary of the PEC material. The numerically calculated dispersion is represented in Fig. 3bi. In this figure, as well as in the other examples throughout the article, the edge state dispersion is shown exclusively within the band gap regions.

The high-frequency band gap, which has a gap Chern number $\mathcal{C}_{\text{gap},2} = +1$ (for $\omega_{\text{c}} = 0.6\omega_{\text{p}}$), supports a single edge state propagating in the $-x$ direction (red), in accordance with the bulk-edge correspondence. The low-frequency gap (which is not topological, $\mathcal{C}_{\text{gap},1} = -0.514$) hosts an edge state that propagates along the $+x$-direction (blue). However, the edge state dispersion does not span the entire gap, i.e., it is not gapless as in topological systems.

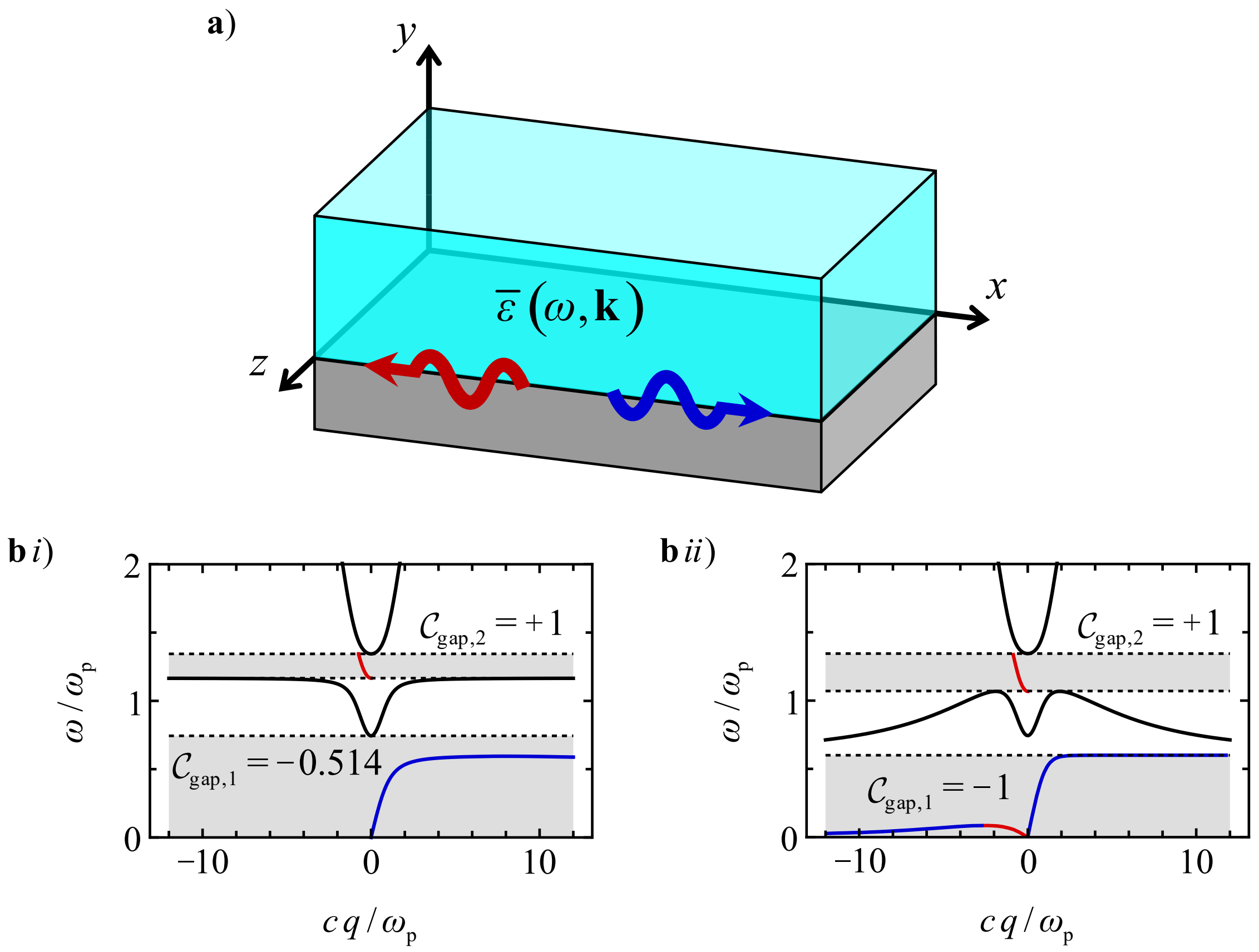

**Fig. 3 (a)** Flat interface ( $y=0$ ) between a PEC ( $y<0$ ) and a magnetized plasma ( $y>0$ ) represented in gray and cyan colors, respectively. For frequencies within the band gaps, the system supports edge states that may propagate in the positive (blue arrow with $\partial\omega/\partial q>0$ ) or negative (red arrow with $\partial\omega/\partial q<0$ ) $x$-direction. **(b)** Edge state dispersion for (i) the local model and (ii) the full-cutoff model with $k_{\max}=5\omega_p/c$. The cyclotron frequency of the magnetized plasma is $\omega_c=0.6\omega_p$. The bulk plasma dispersion is represented with solid black lines, while the dispersion of the edge states propagating forward/backward is represented with blue/red colors, respectively. The band gaps are shaded in a light gray color.

For the full-cutoff model, the system is topological, and the lower-frequency gap is characterized by the gap Chern number $\mathcal{C}_{gap,1}=-1$. As the material response is spatially dispersive for this model, the characteristic equation (21) supports now $N=3$ independent solutions. Consequently, the characterization of the edge states involves the use of two scalar additional boundary conditions (ABCs) [51], besides the standard PEC boundary condition ($E_x = 0$). As detailed in Appendix B, these ABCs can be expressed in a vector form as $\mathbf{J}|_{y=0^+} = \mathbf{0}$. It can be shown that this vector ABC ensures the conservation of the power flow through the spatially dispersive material and the conservation of energy.

The numerically calculated edge state dispersion is represented in Fig. 3bii. Curiously, the number of edge states varies between one and three depending on the frequency of operation. However, throughout the entire gap, there is always one extra mode propagating forward, in agreement with the bulk-edge correspondence ( $N_+ - N_- = 1$). Note that the branch with $q < 0$ crosses the bulk dispersion (static-like $\omega = 0$ band) both at $q = 0^-$ and $q = -\infty$. This branch comprises a mode propagating along $+x$ and another mode propagating along $-x$. On the other hand, the branch with $q > 0$ spans the entire gap and crosses the bulk dispersion at $q = 0^+$ and $q = +\infty$.

The wavevector cutoff can be implemented in practice by introducing an air gap of width $d \sim 1/k_{\max}$ between the magnetized plasma and the PEC regions as explained in Refs. [51, 56, 57]. This property is consistent with the fact that the ABC for the spatially dispersive model ($\mathbf{J}|_{y=0^+} = \mathbf{0}$) ensures the vanishing of the nonreciprocal current density in the material near the PEC interface.

*B. Hydrodynamic Model*

For the hydrodynamic model, the bulk magnetized plasma has a single band gap with a trivial gap Chern number $\mathcal{C}_{\rm gap}=0$. The characteristic equation (21) supports $N=2$ solutions. The edge state dispersion is found by imposing the usual PEC boundary condition ($\hat{\mathbf{x}}\cdot\mathbf{E}\big|_{y=0^+}=0$) and, in addition, the vanishing of the normal component of the electric current density ($\hat{\mathbf{y}}\cdot\mathbf{J}\big|_{y=0^+}=0$) [64]. The details can be found in Appendix B. In this case, as illustrated in Fig. 4a, the system supports exactly two gapless counter-propagating edge states [66, 67]. Therefore, the net number of unidirectional states is zero, exactly as predicted by the bulk-edge correspondence.

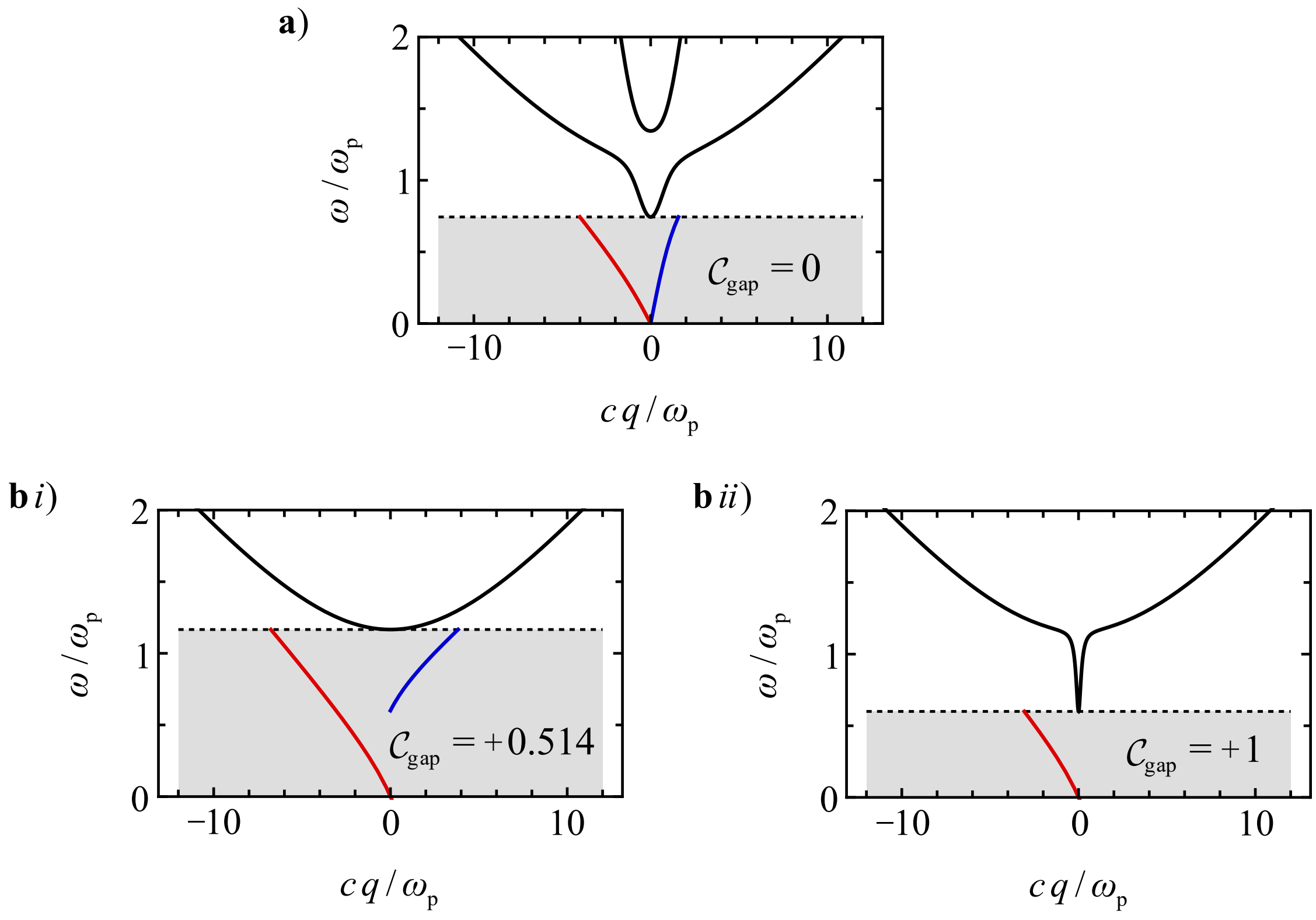

**Fig. 4** Edge state dispersion for the **(a)** exact and **(b)** quasi-static formulations of the hydrodynamic model. For the quasi-static models, bi) represents a system without a low-frequency cutoff, whereas ii) represents a regularized system with $k_{\rm min}=0.2\omega_{\rm p}/c$. The geometry is the same as in Fig. 3a and the magnetized

plasma is characterized by the parameters $\omega_{\mathrm{c}} = 0.6\omega_{\mathrm{p}}$, $\beta = 0.15c$. The bulk plasma dispersion is represented with solid black lines, while the dispersion of the edge states propagating forward/backward is represented with blue/red colors, respectively. The band gaps are shaded in a light gray color.

### *C. Quasi-Static Limit*

Similar to the exact hydrodynamic model, both quasi-static models yield $N = 2$ solutions for the characteristic equation (21). The edge states are found by imposing the boundary conditions $\phi\big|_{y=0^+} = 0$ and $\hat{\mathbf{y}}\cdot\mathbf{J}\big|_{y=0^+} = 0$. The first boundary condition ensures that the tangential electric field vanishes at the PEC boundary.

We begin by analyzing the model with an ill-defined topology, i.e., without the low-spatial frequency cutoff ($k_{\mathrm{min}} = 0$). Akin to the exact hydrodynamic model, the edge state dispersion comprises two counter-propagating modes (see Fig. 4bi). However, one of the edge states (blue curve) is not gapless. This is attributed to the non-integer nature of the gap Chern number ($\mathcal{C}_{\mathrm{gap}} = +0.514$).

For $0 < k_{\mathrm{min}} < \infty$, the quasi-static model becomes topological, resulting in a gap Chern number $\mathcal{C}_{\mathrm{gap}} = +1$. Now, in agreement with the bulk-edge correspondence, the interface supports a single edge-state propagating along the $-x$ direction (Fig. 4bii). It is relevant to note that in this model the band gap width is highly dependent on the cutoff value $k_{\mathrm{min}}$. Specifically, a finite $k_{\mathrm{min}}$ results in a gap width that effectively shifts the mode propagating towards the $+x$ direction in Fig. 4bi outside the band gap in Fig. 4bii.

## V. Discussion and Conclusions

The most striking feature of a Chern insulator is that the topological nature of its edge-type channels makes them robust against weak perturbations, i.e., continuous deformations that do not close the relevant band gap. For example, minor fabrication errors may alter the exact edge dispersion but the asymmetry in the number of modes propagating in each direction remains the same.

Therefore, it may appear puzzling that arbitrarily weak non-local perturbations can give rise to so disparate edge-state dispersions, as summarized in Table 1. To understand this, it is essential to recognize that the original system, i.e., the local magnetized plasma, is topologically ill-defined. For this reason, any perturbation that regularizes its behavior for large wavenumbers, no matter how small it is, abruptly changes the gap Chern number to an integer value. This in turn has important implications for the edge state dispersion. A geometrical illustration of this property was provided in Refs. [57, 58]: a torus with a vanishing inner radius has an ill-defined topology due to the cusp at its center, which makes it a non-differentiable surface. An infinitesimal deformation of the torus either opens or closes the hole, altering the genus of the surface in a discontinuous and non-unique manner.

**Table 1** – Overview of the gap Chern numbers and number of edge states for the lower-frequency gap and different models of a magnetized plasma with $\omega_c = 0.6\omega_p$. The magnetized plasma is surrounded by a PEC wall.

| **Model** | $\mathcal{C}_{\text{gap}}$ | **Number of Edge States** | |
|---|---|---|---|
| | | $N_-$ | $N_+$ |
| Local | –0.514 | 0 | 0/1 |
| Full-Cutoff | –1 | 0/1 | 1/2 |
| Hydrodynamic | 0 | 1 | 1 |
| Quasi-Static | +0.514 | 1 | 0/1 |
| Quasi-Static with Cutoff | +1 | 1 | 0 |

To better understand how the regularization of the topology impacts the bulk dispersions for arbitrarily weak perturbations, let us focus on the full-cutoff regularization and on the hydrodynamic model. Figures 5a and 5b show that the bulk dispersions coincide with that of the local model (black lines) for wavenumbers satisfying $k << k_{\text{max}}$ and $k << \omega_{\text{p}} / \beta$. For sufficiently weak perturbations, it is impossible or extremely challenging to experimentally probe the nonlocal effects of these media within a realistic range of wavenumbers. Therefore, it is often reasonable to dismiss them in most practical applications. However, the gap Chern number is a global property of the system, and the asymptotic behavior of the material response plays a crucial role in it, leading to distinct classes of Chern insulators with very different edge dispersions, as shown in Sect. IV.

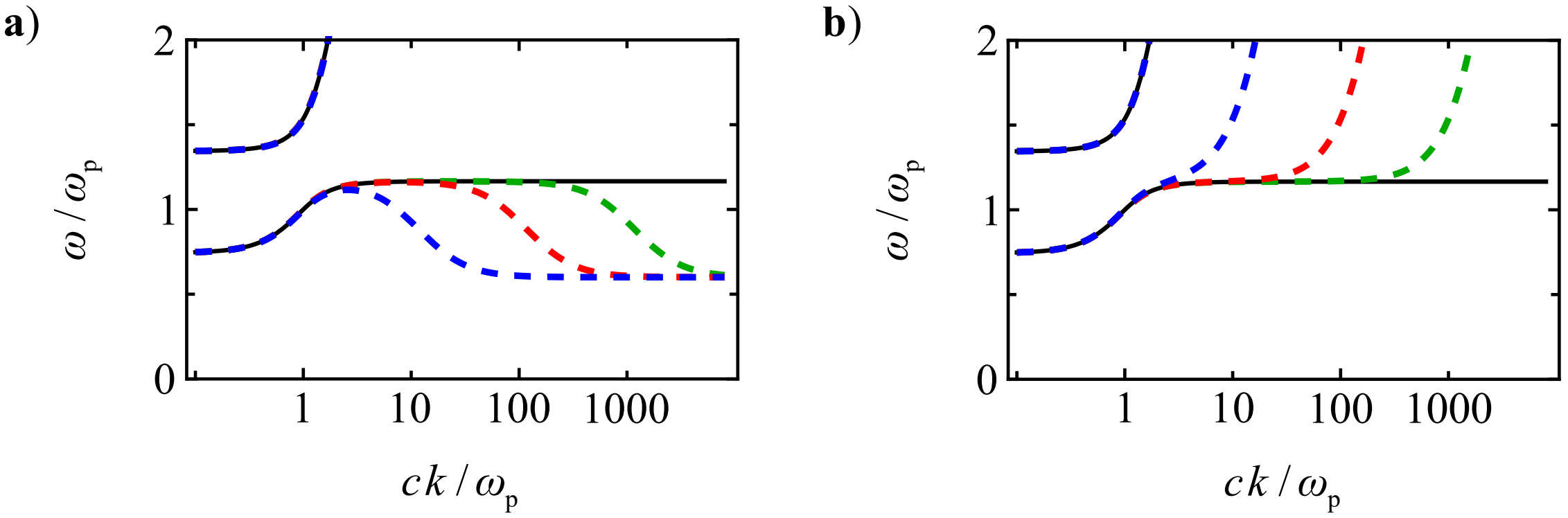

**Fig. 5** Comparison between the TM band structures of a bulk magnetized plasma with a local model (solid black lines) and two regularized models: **(a)** spatial full-cutoff model with $k_{\text{max}} = 10\omega_{\text{p}} / c$ (dashed blue), $k_{\text{max}} = 100\omega_{\text{p}} / c$ (dashed red) and $k_{\text{max}} = 1000\omega_{\text{p}} / c$ (dashed green); **(b)** hydrodynamic model for $\beta = 0.1c$ (dashed blue), $\beta = 0.01c$ (dashed red) and $\beta = 0.001c$ (dashed green).

In summary, we have explored different physical models to describe a continuous magnetized plasma: a local model, a full-cutoff model, a hydrodynamic model, and a quasi-static model. In the absence of nonlocal effects, the system is topologically ill-defined. In agreement with previous works, we have shown that due to this property the edge dispersion is not gapless. In other words, the number of modes propagating in each direction is not constant and depends on the frequency of operation within the band gap.

Furthermore, we have shown that arbitrarily small nonlocal perturbations can lead to different classes of well-defined Chern insulators. This discontinuous behavior naturally leads to very distinct edge dispersions, as consistently predicted by the bulk-edge correspondence. In topological systems, the edge state dispersions are such that the net number of unidirectional modes across the entire bandgap is constant. Our findings underscore the importance of the high-spatial frequency response of a material when dealing with topology-related phenomena: topological invariants are determined by the global properties of the systems, regardless of how similar the systems may appear in the low-spatial frequency region of the parametric space.

**Acknowledgements:** This work was partially funded by the Institution of Engineering and Technology (IET), by the Simons Foundation (Award 733700) and by Instituto de Telecomunicações under Project No. UIDB/50008/2020.

## Appendix A: Dispersion relations, Berry potentials and Chern numbers of the bulk magnetized plasma

In this Appendix, we present the dispersion relations, Berry potentials and Chern numbers of the bulk magnetized plasma for the different physical models discussed in the main text.

### A. *Local and Full-cutoff Models*

The dispersion equation for the full-cutoff model is,

$$\det\left[H_{\text{cutoff}}\left(\mathbf{k}\right)-\omega_{n,\mathbf{k}}\mathbf{1}\right]=0, \qquad \text{(A1)}$$

where $H_{\text{cutoff}}(\mathbf{k})$ is defined as in Eq. (12) of the main text. The dispersion of the TM waves is governed by five frequency bands:

$$\frac{\omega_{0,\mathbf{k}}}{\omega_{\text{p}}}=0, \tag{A2a}$$

$$\frac{\omega_{\pm1,\mathbf{k}}}{\omega_{\text{p}}}=\pm\sqrt{\frac{1}{1+k^2/k_{\max}^2}+\frac{c^2k^2+\omega_{\text{c}}^2}{2\omega_{\text{p}}^2}-\sqrt{\frac{\left(c^2k^2-\omega_{\text{c}}^2\right)^2}{4\omega_{\text{p}}^4}+\frac{\omega_{\text{c}}^2/\omega_{\text{p}}^2}{1+k^2/k_{\max}^2}}}, \tag{A2b}$$

$$\frac{\omega_{\pm2,\mathbf{k}}}{\omega_{\text{p}}}=\pm\sqrt{\frac{1}{1+k^2/k_{\max}^2}+\frac{c^2k^2+\omega_{\text{c}}^2}{2\omega_{\text{p}}^2}+\sqrt{\frac{\left(c^2k^2-\omega_{\text{c}}^2\right)^2}{4\omega_{\text{p}}^4}+\frac{\omega_{\text{c}}^2/\omega_{\text{p}}^2}{1+k^2/k_{\max}^2}}}. \tag{A2c}$$

The corresponding normalized eigenfunctions are:

$$\boldsymbol{\psi}_{n,\mathbf{k}}=\frac{1}{C_{n,\mathbf{k}}}\begin{pmatrix} -i\dfrac{c\omega_{\text{c}}/\omega_{\text{p}}^2}{1+k^2/k_{\max}^2}\mathbf{k}-\dfrac{c\omega_{n,\mathbf{k}}}{\omega_{\text{p}}^2}\left(\dfrac{1}{1+k^2/k_{\max}^2}-\dfrac{\omega_{n,\mathbf{k}}^2-\omega_{\text{c}}^2}{\omega_{\text{p}}^2}\right)\hat{\mathbf{z}}\times\mathbf{k} \\ \left[\left(\dfrac{\omega_{n,\mathbf{k}}^2}{\omega_{\text{p}}^2}-\dfrac{1}{1+k^2/k_{\max}^2}\right)^2-\dfrac{\omega_{n,\mathbf{k}}^2\omega_{\text{c}}^2}{\omega_{\text{p}}^4}\right]\hat{\mathbf{z}} \\ \sqrt{\dfrac{1}{1+k^2/k_{\max}^2}}\left[\dfrac{c\omega_{n,\mathbf{k}}\omega_{\text{c}}}{\omega_{\text{p}}^3}\mathbf{k}-i\dfrac{c}{\omega_{\text{p}}}\left(\dfrac{1}{1+k^2/k_{\max}^2}-\dfrac{\omega_{n,\mathbf{k}}^2}{\omega_{\text{p}}^2}\right)\hat{\mathbf{z}}\times\mathbf{k}\right] \end{pmatrix}. \tag{A3}$$

with

$$\begin{aligned} C_{n,\mathbf{k}}^2 &= \frac{c^2k^2/\omega_{\text{p}}^2}{1+k^2/k_{\max}^2}\left(\frac{\omega_{n,\mathbf{k}}^2\omega_{\text{c}}^2}{\omega_{\text{p}}^4}+\left(\frac{1}{1+k^2/k_{\max}^2}-\frac{\omega_{n,\mathbf{k}}^2}{\omega_{\text{p}}^2}\right)^2\right) \\ &+\frac{c^2k^2}{\omega_{\text{p}}^2}\left[\frac{\omega_{\text{c}}^2}{\omega_{\text{p}}^2}\left(\frac{1}{1+k^2/k_{\max}^2}\right)^2+\frac{\omega_{n,\mathbf{k}}^2}{\omega_{\text{p}}^2}\left(\frac{1}{1+k^2/k_{\max}^2}-\frac{\omega_{n,\mathbf{k}}^2-\omega_{\text{c}}^2}{\omega_{\text{p}}^2}\right)^2\right] \\ &+\left[\left(\frac{\omega_{n,\mathbf{k}}^2}{\omega_{\text{p}}^2}-\frac{1}{1+k^2/k_{\max}^2}\right)^2-\frac{\omega_{n,\mathbf{k}}^2\omega_{\text{c}}^2}{\omega_{\text{p}}^4}\right]^2. \end{aligned} \tag{A4}$$

Each band $\omega_{n,\mathbf{k}}$ is characterized by the Berry potential

$$\mathbf{A}_{n,\mathbf{k}} = \frac{-4}{C_{n,\mathbf{k}}^2} \frac{c^2 \omega_{\mathrm{c}} \omega_{n,\mathbf{k}} / \omega_{\mathrm{p}}^4}{1 + k^2 / k_{\max}^2} \left( \frac{1}{1 + k^2 / k_{\max}^2} - \frac{\omega_{n,\mathbf{k}}^2}{\omega_{\mathrm{p}}^2} + \frac{\omega_{\mathrm{c}}^2}{2\omega_{\mathrm{p}}^2} \right) \hat{\mathbf{z}} \times \mathbf{k}, \tag{A5}$$

where we have used $\mathbf{A}_{n,\mathbf{k}} = i \boldsymbol{\psi}_{n,\mathbf{k}}^* \cdot \nabla_{\mathbf{k}} \boldsymbol{\psi}_{n,\mathbf{k}}$. The formulas for the local model are obtained from the previous ones by setting $k_{\max} = \infty$.

In the local case ($k_{\max} = \infty$), the (non-static) bands have the following Chern numbers:

$$\begin{aligned} \mathcal{C}_{\pm 1} &= \lim_{k \to +\infty} \left( \hat{\mathbf{z}} \times \mathbf{k} \right) \cdot \mathbf{A}_{\pm 1,\mathbf{k}} - \lim_{k \to 0} \left( \hat{\mathbf{z}} \times \mathbf{k} \right) \cdot \mathbf{A}_{\pm 1,\mathbf{k}} \\ &= \pm \frac{1}{\sqrt{1 + \omega_{\mathrm{p}}^2 / \omega_{\mathrm{c}}^2}} - \left( \mp \operatorname{sgn}\left[\omega_{\mathrm{c}}\right] \right) = \pm \operatorname{sgn}\left[\omega_{\mathrm{c}}\right] \left( 1 + \frac{1}{\sqrt{1 + \omega_{\mathrm{p}}^2 / \omega_{\mathrm{c}}^2}} \right), \end{aligned} \tag{A6a}$$

$$\begin{aligned} \mathcal{C}_{\pm 2} &= \lim_{\|\mathbf{k}\| \to +\infty} \left( \hat{\mathbf{z}} \times \mathbf{k} \right) \cdot \mathbf{A}_{\pm 2,\mathbf{k}} - \lim_{\|\mathbf{k}\| \to 0} \left( \hat{\mathbf{z}} \times \mathbf{k} \right) \cdot \mathbf{A}_{\pm 2,\mathbf{k}} \\ &= 0 - \left( \pm \operatorname{sgn}\left[\omega_{\mathrm{c}}\right] \right) = \mp \operatorname{sgn}\left[\omega_{\mathrm{c}}\right]. \end{aligned} \tag{A6b}$$

When the wavevector cutoff $k_{\max} > 0$ is finite, the topological phases become regularized:

$$\begin{aligned} \mathcal{C}_{\pm 1} &= \lim_{k \to +\infty} \left( \hat{\mathbf{z}} \times \mathbf{k} \right) \cdot \mathbf{A}_{\pm 1,\mathbf{k}} - \lim_{k \to 0} \left( \hat{\mathbf{z}} \times \mathbf{k} \right) \cdot \mathbf{A}_{\pm 1,\mathbf{k}} \\ &= \pm \operatorname{sgn}\left[\omega_{\mathrm{c}}\right] - \left( \mp \operatorname{sgn}\left[\omega_{\mathrm{c}}\right] \right) = \pm 2 \operatorname{sgn}\left[\omega_{\mathrm{c}}\right], \end{aligned} \tag{A7a}$$

$$\begin{aligned} \mathcal{C}_{\pm 2} &= \lim_{k \to +\infty} \left( \hat{\mathbf{z}} \times \mathbf{k} \right) \cdot \mathbf{A}_{\pm 2,\mathbf{k}} - \lim_{k \to 0} \left( \hat{\mathbf{z}} \times \mathbf{k} \right) \cdot \mathbf{A}_{\pm 2,\mathbf{k}} \\ &= 0 - \left( \pm \operatorname{sgn}\left[\omega_{\mathrm{c}}\right] \right) = \mp \operatorname{sgn}\left[\omega_{\mathrm{c}}\right]. \end{aligned} \tag{A7b}$$

### *B. Hydrodynamic Model*

The hydrodynamic model [see Eq. (15)] is characterized by four non-trivial bands with dispersions:

$$\frac{\omega_{\pm 1,\mathbf{k}}}{\omega_{\mathrm{p}}} = \pm\sqrt{1+\frac{\left(c^2+\beta^2\right)k^2+\omega_{\mathrm{c}}^2}{2\omega_{\mathrm{p}}^2} - \sqrt{\frac{\left(\left(c^2-\beta^2\right)k^2-\omega_{\mathrm{c}}^2\right)^2}{4\omega_{\mathrm{p}}^4}+\frac{\omega_{\mathrm{c}}^2}{\omega_{\mathrm{p}}^2}}}, \tag{A8a}$$

$$\frac{\omega_{\pm 2,\mathbf{k}}}{\omega_{\mathrm{p}}} = \pm\sqrt{1+\frac{\left(c^2+\beta^2\right)k^2+\omega_{\mathrm{c}}^2}{2\omega_{\mathrm{p}}^2} + \sqrt{\frac{\left(\left(c^2-\beta^2\right)k^2-\omega_{\mathrm{c}}^2\right)^2}{4\omega_{\mathrm{p}}^4}+\frac{\omega_{\mathrm{c}}^2}{\omega_{\mathrm{p}}^2}}}. \tag{A8b}$$

The corresponding (normalized) eigenvectors are given by:

$$\boldsymbol{\psi}_{n,\mathbf{k}} = \frac{1}{C_{n,\mathbf{k}}}\begin{pmatrix} -i\frac{c\omega_{\mathrm{c}}}{\omega_{\mathrm{p}}^2}\mathbf{k} - \frac{c\omega_{n,\mathbf{k}}}{\omega_{\mathrm{p}}^2}\left(1-\frac{\omega_{n,\mathbf{k}}^2-\omega_{\mathrm{c}}^2-\beta^2k^2}{\omega_{\mathrm{p}}^2}\right)\hat{\mathbf{z}}\times\mathbf{k} \\ \left[\left(1+\frac{\beta^2k^2-\omega_{n,\mathbf{k}}^2}{\omega_{\mathrm{p}}^2}\right)\left(1-\frac{\omega_{n,\mathbf{k}}^2}{\omega_{\mathrm{p}}^2}\right)-\frac{\omega_{\mathrm{c}}^2\omega_{n,\mathbf{k}}^2}{\omega_{\mathrm{p}}^4}\right]\hat{\mathbf{z}} \\ \frac{\omega_{\mathrm{c}}\beta c k^2}{\omega_{\mathrm{p}}^3} \\ \frac{c\omega_{n,\mathbf{k}}\omega_{\mathrm{c}}}{\omega_{\mathrm{p}}^3}\mathbf{k} - i\frac{c}{\omega_{\mathrm{p}}}\left(1+\frac{\beta^2k^2-\omega_{n,\mathbf{k}}^2}{\omega_{\mathrm{p}}^2}\right)\hat{\mathbf{z}}\times\mathbf{k} \end{pmatrix}. \tag{A9}$$

with

$$\begin{aligned} C_{n,\mathbf{k}}^2 &= \frac{c^2k^2}{\omega_{\mathrm{p}}^2}\left[\frac{\omega_{n,\mathbf{k}}^2}{\omega_{\mathrm{p}}^2}\left(1-\frac{\omega_{n,\mathbf{k}}^2-\omega_{\mathrm{c}}^2-\beta^2k^2}{\omega_{\mathrm{p}}^2}\right)^2+\left(1+\frac{\beta^2k^2-\omega_{n,\mathbf{k}}^2}{\omega_{\mathrm{p}}^2}\right)^2\right] \\ &+\frac{c^2k^2\omega_{\mathrm{c}}^2}{\omega_{\mathrm{p}}^4}\left(1+\frac{\omega_{n,\mathbf{k}}^2}{\omega_{\mathrm{p}}^2}+\frac{\beta^2k^2}{\omega_{\mathrm{p}}^2}\right)+\left[\left(1+\frac{\beta^2k^2-\omega_{n,\mathbf{k}}^2}{\omega_{\mathrm{p}}^2}\right)\left(1-\frac{\omega_{n,\mathbf{k}}^2}{\omega_{\mathrm{p}}^2}\right)-\frac{\omega_{\mathrm{c}}^2\omega_{n,\mathbf{k}}^2}{\omega_{\mathrm{p}}^4}\right]^2. \end{aligned} \tag{A10}$$

The Berry potential is given by

$$\mathbf{A}_{n,\mathbf{k}} = -4\frac{c^2\omega_{n,\mathbf{k}}\omega_{\mathrm{c}}}{\omega_{\mathrm{p}}^4C_{n,\mathbf{k}}^2}\left[1+\frac{\beta^2k^2-\omega_{n,\mathbf{k}}^2}{\omega_{\mathrm{p}}^2}+\frac{\omega_{\mathrm{c}}^2}{2\omega_{\mathrm{p}}^2}\right]\hat{\mathbf{z}}\times\mathbf{k}. \tag{A11}$$

The bands have the following Chern numbers:

$$\begin{aligned} \mathcal{C}_{\pm 1} &= \lim_{k\to+\infty}\left(\hat{\mathbf{z}}\times\mathbf{k}\right)\cdot\mathbf{A}_{\pm 1,\mathbf{k}} - \lim_{k\to 0}\left(\hat{\mathbf{z}}\times\mathbf{k}\right)\cdot\mathbf{A}_{\pm 1,\mathbf{k}} \\ &= 0-\left(\mp\operatorname{sgn}\left[\omega_{\mathrm{c}}\right]\right) = \pm\operatorname{sgn}\left[\omega_{\mathrm{c}}\right], \end{aligned} \tag{A12a}$$

$$\mathcal{C}_{\pm 2} = \lim_{k\to+\infty} (\hat{\mathbf{z}}\times\mathbf{k})\cdot\mathbf{A}_{\pm 2,\mathbf{k}} - \lim_{k\to 0} (\hat{\mathbf{z}}\times\mathbf{k})\cdot\mathbf{A}_{\pm 2,\mathbf{k}} \\ = 0-\left(\pm\operatorname{sgn}[\omega_c]\right) = \mp\operatorname{sgn}[\omega_c]. \tag{A12b}$$

## C. Quasi-Static Limit

In the quasi-static limit, the spectral problem associated with the hydrodynamic model reduces to $H_{QS}^{\text{cutoff}}(\mathbf{k})\cdot\boldsymbol{\psi}_{n,\mathbf{k}} = \frac{\omega_{n,\mathbf{k}}}{\omega_p}\boldsymbol{\psi}_{n,\mathbf{k}}$, with $H_{QS}^{\text{cutoff}}(\mathbf{k})$ defined as in Eq. (19). It yields two non-trivial bands with the following dispersions:

$$\frac{\omega_{\pm 1,\mathbf{k}}}{\omega_p} = \pm\sqrt{\frac{\omega_c^2}{\omega_p^2} + \left(\beta^2 + \frac{\omega_p^2}{k^2+k_{\min}^2}\right)\frac{k^2}{\omega_p^2}}\,. \tag{A13}$$

The corresponding (normalized) eigenvectors are given by:

$$\boldsymbol{\psi}_{n,\mathbf{k}} = \frac{1}{\sqrt{k^2\left(\beta^2+\frac{\omega_p^2}{k^2+k_{\min}^2}\right)\left(\omega_{n,\mathbf{k}}^2+\omega_c^2\right) + \left(\omega_{n,\mathbf{k}}^2-\omega_c^2\right)^2}} \begin{pmatrix} \omega_{n,\mathbf{k}}^2 - \omega_c^2 \\ \sqrt{\beta^2+\frac{\omega_p^2}{k^2+k_{\min}^2}}\left(\omega_{n,\mathbf{k}}\mathbf{k} + i\omega_c\hat{\mathbf{z}}\times\mathbf{k}\right) \end{pmatrix}. \tag{A14}$$

The Berry potential satisfies

$$\mathbf{A}_{n,\mathbf{k}} = \frac{2\left(\beta^2+\frac{\omega_p^2}{k^2+k_{\min}^2}\right)\omega_c\omega_{n,\mathbf{k}}\hat{\mathbf{z}}\times\mathbf{k}}{k^2\left(\beta^2+\frac{\omega_p^2}{k^2+k_{\min}^2}\right)\left(\omega_{n,\mathbf{k}}^2+\omega_c^2\right) + \left(\omega_{n,\mathbf{k}}^2-\omega_c^2\right)^2}\,. \tag{A15}$$

Without the low-wavevector cutoff ($k_{\min}=0$), the bands have the following Chern numbers:

$$\mathcal{C}_{\pm 1} = \lim_{k\to+\infty} (\hat{\mathbf{z}}\times\mathbf{k})\cdot\mathbf{A}_{\pm 1,\mathbf{k}} - \lim_{k\to 0} (\hat{\mathbf{z}}\times\mathbf{k})\cdot\mathbf{A}_{\pm 1,\mathbf{k}} \\ = 0-\left(\pm\operatorname{sgn}[\omega_c]\frac{1}{\sqrt{1+\omega_p^2/\omega_c^2}}\right) = \mp\operatorname{sgn}[\omega_c]\frac{1}{\sqrt{1+\omega_p^2/\omega_c^2}}. \tag{A16}$$

With the wavevector cutoff ( $k_{\min} > 0$ ), we find

$$\begin{aligned} \mathcal{C}_{\pm 1} &= \lim_{k\to+\infty} \left(\hat{\mathbf{z}}\times\mathbf{k}\right)\cdot\mathbf{A}_{\pm1,\mathbf{k}} - \lim_{k\to 0} \left(\hat{\mathbf{z}}\times\mathbf{k}\right)\cdot\mathbf{A}_{\pm1,\mathbf{k}} \\ &= 0-\left(\pm\,\mathrm{sgn}\left[\omega_{\mathrm{c}}\right]\right) = \mp\,\mathrm{sgn}\left[\omega_{\mathrm{c}}\right]. \end{aligned} \tag{A17}$$

Note that the above result remains true when $\omega_{\mathrm{p}} = 0$, as in Ref. [65], which is a particular case insensitive to the cut-off.

## Appendix B: Derivation of the edge states dispersion

In this Appendix, we deduce the dispersion relations for TM edge states propagating at the interface $y=0$ between a magnetized plasma ( $y>0$ ) and a PEC ( $y<0$ ) represented in Fig. 3a.

*A. Local and Full-cutoff Models*

For a local magnetized plasma [Eq. (9)], the characteristic equation (21) reduces to

$$\frac{c^2k^2}{\omega_{\mathrm{p}}^2} = \frac{\omega^2}{\omega_{\mathrm{p}}^2} - \frac{\omega^2-\omega_{\mathrm{p}}^2}{\omega^2-\omega_{\mathrm{p}}^2-\omega_{\mathrm{c}}^2}. \tag{B1}$$

with $k^2 = q^2 - \gamma^2$. It yields exactly a single solution for $\gamma > 0$ ( $N=1$). The edge state dispersion is found by enforcing $\hat{\mathbf{x}}\cdot\mathbf{E}\big|_{y=0^+} = 0$ in Eq. (A3) with $\mathbf{k} = q\hat{\mathbf{x}} + i\gamma\hat{\mathbf{y}}$ and $k_{\max} = \infty$. This yields the well-known result:

$$\omega_{\mathrm{c}} q - \omega\left(1-\frac{\omega^2-\omega_{\mathrm{c}}^2}{\omega_{\mathrm{p}}^2}\right)\gamma = 0\,. \tag{B2}$$

For the full-cutoff model [Eq. (12)], the characteristic equation (21) yields

$$\frac{\left(\omega^2-\omega_{\mathrm{c}}^2\right)\left(c^2k^2-\omega^2\right)}{\omega_{\mathrm{p}}^4} - \frac{k_{\max}^2}{k^2+k_{\max}^2}\left(\frac{k_{\max}^2}{k^2+k_{\max}^2} + \frac{c^2k^2-2\omega^2}{\omega_{\mathrm{p}}^2}\right) = 0. \tag{B3}$$

This equation supports $N=3$ positive solutions, $\gamma_1$, $\gamma_2$ and $\gamma_3$, for a fixed frequency $\omega$. Using again Eqs. (A3) and (22), and imposing the boundary conditions $\hat{\mathbf{x}}\cdot\mathbf{E}\big|_{y=0^+}=0$ and $\mathbf{J}\big|_{y=0^+}=\mathbf{0}$, we find the edge states dispersion equation takes the form:

$$\det\begin{pmatrix} a_1 & a_2 & a_3 \\ b_1 & b_2 & b_3 \\ c_1 & c_2 & c_3 \end{pmatrix}=0 \tag{B4}$$

with

$$\begin{aligned} a_n &= \frac{\omega_{\rm c} q}{1+k_n^2/k_{\max}^2}-\left(\frac{1}{1+k_n^2/k_{\max}^2}-\frac{\omega^2-\omega_{\rm c}^2}{\omega_{\rm p}^2}\right)\omega\gamma_n, \\ b_n &= \frac{1}{1+k_n^2/k_{\max}^2}\left[\frac{\omega\omega_{\rm c}}{\omega_{\rm p}^2}q-\left(\frac{1}{1+k_n^2/k_{\max}^2}-\frac{\omega^2}{\omega_{\rm p}^2}\right)\gamma_n\right], \\ c_n &= \frac{1}{1+k_n^2/k_{\max}^2}\left[\frac{\omega\omega_{\rm c}}{\omega_{\rm p}^2}\gamma_n-\left(\frac{1}{1+k_n^2/k_{\max}^2}-\frac{\omega^2}{\omega_{\rm p}^2}\right)q\right]. \end{aligned} \tag{B5}$$

To simplify the notations, we introduced $k_n^2=q^2-\gamma_n^2, \quad n=1,2,3$.

It is relevant to note that the full-cutoff model stems from the "regularized" transport equation,

$$+i\varepsilon_0\omega_{\rm p}^2\mathbf{E}=\left[\omega\mathbf{1}-i\omega_{\rm c}\hat{\mathbf{z}}\times\mathbf{1}\right]\cdot\left(1-\frac{\nabla^2}{k_{\max}^2}\right)\mathbf{J}. \tag{B6}$$

It can be shown that the ABC $\mathbf{J}\big|_{y=0^+}=\mathbf{0}$ ensures the conservation of power flow at the interface of the spatially dispersive material with another conventional (local) dielectric or metal.

*B. Hydrodynamic Model*

For the hydrodynamic model [Eq. (15)], the characteristic equation (21) reduces to:

$$\frac{\left(\omega^2-\omega_{\rm p}^2-c^2k^2\right)\left(\omega^2-\omega_{\rm p}^2-\beta^2k^2\right)}{\omega_{\rm p}^4}=\frac{\omega_{\rm c}^2\left(\omega^2-c^2k^2\right)}{\omega_{\rm p}^4} \tag{B7}$$

with $k^2=q^2-\gamma^2$. Now, there are $N=2$ solutions with $\gamma>0$, which we denote by $\gamma_1$ and $\gamma_2$.

The state vector associated with the edge states is obtained from Eq. (22) with the help of Eq. (A9). By imposing the boundary conditions $\hat{\mathbf{x}}\cdot\mathbf{E}\big|_{y=0^+}=0$ and $\hat{\mathbf{y}}\cdot\mathbf{J}\big|_{y=0^+}=0$, one finds that the unknown coefficients $\alpha_1$ and $\alpha_2$ must satisfy the following linear system:

$$\bar{M}\cdot\begin{pmatrix}\alpha_1\\ \alpha_2\end{pmatrix}=\begin{pmatrix}0\\ 0\end{pmatrix} \tag{B8}$$

with

$$\bar{M}=\begin{pmatrix}\frac{\omega_{\rm c}}{\omega_{\rm p}}q-\left(1-\frac{\omega^2-\omega_{\rm c}^2-\beta^2k_1^2}{\omega_{\rm p}^2}\right)\frac{\omega}{\omega_{\rm p}}\gamma_1 & \frac{\omega_{\rm c}}{\omega_{\rm p}}q-\left(1-\frac{\omega^2-\omega_{\rm c}^2-\beta^2k_2^2}{\omega_{\rm p}^2}\right)\frac{\omega}{\omega_{\rm p}}\gamma_2\\ \frac{\omega\omega_{\rm c}}{\omega_{\rm p}^2}\gamma_1-\left(1+\frac{\beta^2k_1^2-\omega^2}{\omega_{\rm p}^2}\right)q & \frac{\omega\omega_{\rm c}}{\omega_{\rm p}^2}\gamma_2-\left(1+\frac{\beta^2k_2^2-\omega^2}{\omega_{\rm p}^2}\right)q\end{pmatrix}. \tag{B9}$$

In the above, we defined $k_n^2=q^2-\gamma_n^2,\quad n=1,2$. The edge state dispersion equation is found by setting $\det\bar{M}=0$.

*C. Quasi-Static Limit*

For the quasi-static formulation of the hydrodynamic model [Eq. (19)], the characteristic equation (21) has two solutions $\gamma_\pm$ defined implicitly by:

$$\frac{\beta^2\gamma_\pm^2}{\omega_\mathrm{p}^2}=\frac{\beta^2 q^2}{\omega_\mathrm{p}^2}-\frac{\omega^2-\omega_\mathrm{p}^2-\omega_\mathrm{c}^2-\beta^2 k_\mathrm{min}^2}{2\omega_\mathrm{p}^2} \pm\sqrt{\frac{\left(\omega^2-\omega_\mathrm{p}^2-\omega_\mathrm{c}^2-\beta^2 k_\mathrm{min}^2\right)^2}{4\omega_\mathrm{p}^4}+\frac{\beta^2 k_\mathrm{min}^2}{\omega_\mathrm{p}^2}\frac{\omega^2-\omega_\mathrm{c}^2}{\omega_\mathrm{p}^2}}. \tag{B10}$$

In the non-regularized case ( $k_\mathrm{min}=0$ ), the above formulas simplify to:

$$\gamma_-=\sqrt{q^2-\frac{\omega^2-\omega_\mathrm{p}^2-\omega_\mathrm{c}^2}{\beta^2}}, \tag{B11a}$$

$$\gamma_+=|q|. \tag{B11b}$$

Using Eq. (A14), the state vector associated with an edge state can be written as (at the origin):

$$\boldsymbol{\psi}=\alpha_+\begin{pmatrix}\omega^2-\omega_\mathrm{c}^2\\ \sqrt{\beta^2+\frac{\omega_\mathrm{p}^2}{k^2+k_\mathrm{min}^2}}\left(\omega\mathbf{k}_++i\omega_\mathrm{c}\hat{\mathbf{z}}\times\mathbf{k}_+\right)\end{pmatrix}+\alpha_-\begin{pmatrix}\omega^2-\omega_\mathrm{c}^2\\ \sqrt{\beta^2+\frac{\omega_\mathrm{p}^2}{k^2+k_\mathrm{min}^2}}\left(\omega\mathbf{k}_-+i\omega_\mathrm{c}\hat{\mathbf{z}}\times\mathbf{k}_-\right)\end{pmatrix}. \tag{B12}$$

with $\mathbf{k}_\pm=q\hat{\mathbf{x}}+i\gamma_\pm\hat{\mathbf{y}}$ . Enforcing the boundary conditions $\phi|_{y=0^+}=0$ and $\hat{\mathbf{y}}\cdot\mathbf{J}|_{y=0^+}=0$, we obtain the following dispersion equation:

$$\left[\omega_\mathrm{p}^2+\beta^2\left(k_+^2+k_\mathrm{min}^2\right)\right]\left(\omega\gamma_++\omega_\mathrm{c}q\right)=\left[\omega_\mathrm{p}^2+\beta^2\left(k_-^2+k_\mathrm{min}^2\right)\right]\left(\omega\gamma_-+\omega_\mathrm{c}q\right). \tag{B13}$$